# Quantum Tunneling Enhanced Hydrogen Desorption from Graphene Surface: Atomic versus Molecular Mechanism

Yangwu Tong（童洋武）[1,2] and Yong Yang（杨勇）[1,2]*

1. Science Island Branch of Graduate School, University of Science and Technology of China, Hefei 230026, China.

2. Key Lab of Photovoltaic and Energy Conservation Materials, Institute of Solid State Physics, HFIPS, Chinese Academy of Sciences, Hefei 230031, China.

We study the desorption mechanism of hydrogen isotopes from graphene surface using first-principles calculations, with focus on the effects of quantum tunneling. At low temperatures, quantum tunneling plays a dominant role in the desorption process of both hydrogen monomers and dimers. In the case of dimer desorption, two types of mechanisms, namely the traditional one-step desorption in the form of molecules (molecular mechanism), and the two-step desorption in the form of individual atoms (atomic mechanism) are studied and compared. For the ortho-dimers, the dominant desorption mechanism is found to switch from the molecular mechanism to the atomic mechanism above a critical temperature, which is respectively ~ 300 K and 200 K for H and D.

*Corresponding Author: yyanglab@issp.ac.cn

The desorption of hydrogen isotopes from graphite/graphene surfaces has garnered significant interest in numerous fields due to its pivotal role in several crucial reactions, including the formation of $H_2$ molecules in the interstellar medium [1], the removal of tritium retention from the first wall of fusion reactors [2-4], the release of hydrogen from graphene-based hydrogen storage materials [5-10], and tailoring the electronic and magnetic properties of graphene [11, 12]. Experimentally, hydrogen can be randomly deposited on graphite surface using standard atomic deposition technique. Previous studies [13] have shown that 80% of the hydrogen atoms are adsorbed on the graphite surface as monomers at very low coverage (0.03%). As the hydrogen coverage of graphite surface increases, hydrogen atoms tend to adsorb on the graphite surface in the form of dimers. For the graphite surface with a hydrogen coverage of 0.15%, the proportion of dimers is more than 95% at 140 K and room temperature [14]. When the hydrogen coverage increases to ~ 1%, the most probable adsorption configurations are still hydrogen dimers [15]. Therefore, desorption of hydrogen dimers is particularly important for hydrogen desorption from graphite surface. However, direct observation of the hydrogen desorption processes from graphite surface in experiments remains challenging. For the desorption of hydrogen dimers, existing theoretical studies mainly focus on the situation of two neighboring hydrogen atoms which leave the surface simultaneously in the form of a hydrogen molecules [16, 17]. However, another possible mechanism, in which one of the hydrogen atoms in the dimer takes the lead in desorption to convert the dimer into a hydrogen monomer, is often overlooked. Meanwhile, another important aspect related to the desorption process of hydrogen is quantum tunneling, which usually plays a nontrivial role in the dynamical processes involving light elements such as hydrogen isotopes. Although the tunneling effect to desorption in the form of molecules has been studied in detail [17], its role in desorption in the form of individual hydrogen atoms has not been thoroughly considered. Furthermore, the possible temperature-induced transition of the dominant desorption mechanism involving quantum tunneling in hydrogen dimers has so far not been investigated. In this work, we carry out a systematic study on the potential barriers and probabilities of hydrogen desorption from the graphene surface for different configurations of hydrogen dimers under two desorption mechanisms, taking into account the isotope effects and quantum tunneling. By comparison with the situation of a classical particle, it is found that quantum tunneling dominates the molecular desorption mechanism at low temperatures and has minor impacts on the two-step atomic mechanism of dimer desorption. A transition of the dominant desorption mechanism of ortho-dimers with H and D as adsorbates occurs at ~ 300 K and 200 K, respectively.

The interactions between H/D and graphene surface are studied using density functional theory (DFT) calculations with DFT-D3 dispersion correction to deal with the van der Waals interactions [18]. The graphene surface is modeled by a (5×5) supercell with a vacuum space of ~ 15 Å in the $z$-direction that repeats periodically along the $xy$ plane. The quantum tunneling of H/D across a given potential field is treated using the transfer matrix (TM) method [19-23]. More details of the theoretical methods can be found in the Supporting Materials.

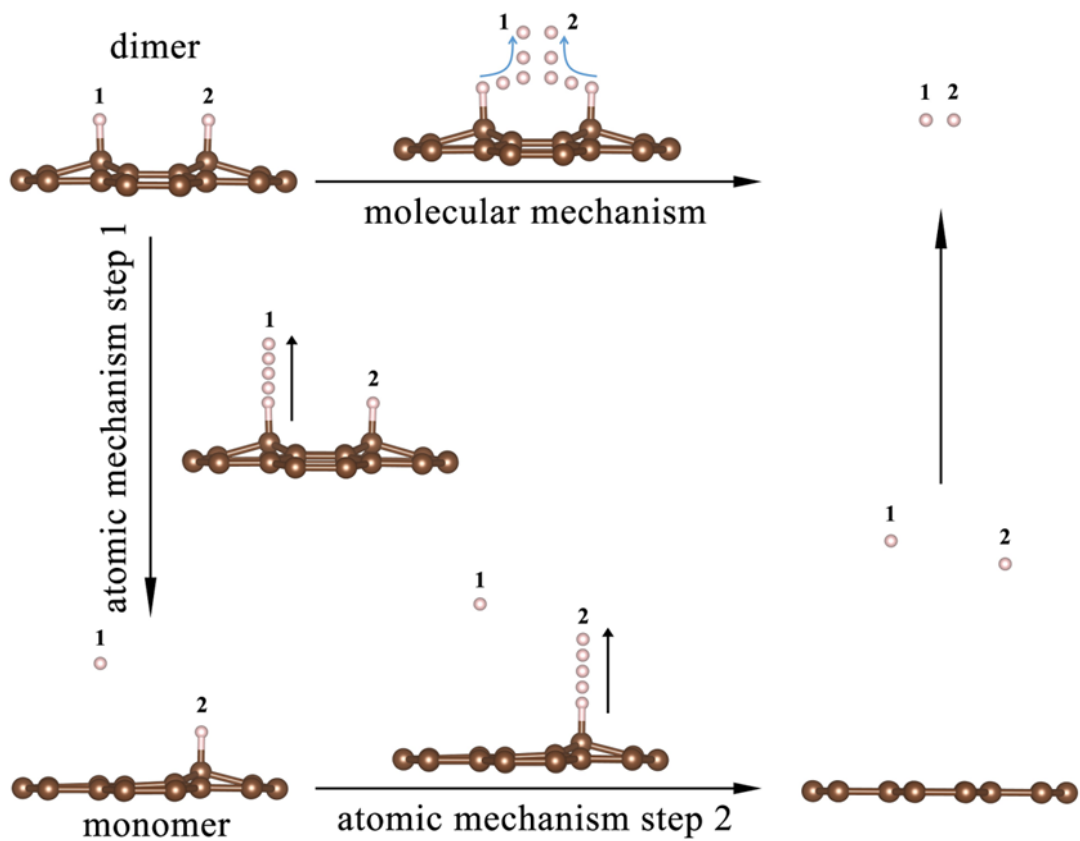

**Fig. 1.** Schematic diagram for the two mechanisms of hydrogen desorption from graphene surface.

The top site of the carbon atom is the only stable chemisorption site for the hydrogen atom on the graphite/graphene surface [24]. For the case where two hydrogen atoms are chemically adsorbed on a hexagonal carbon ring, there are three different types of dimer structures, which are ortho-dimer, meta-dimer, and para-dimer (see Fig. S1). As seen from Fig. S1, when hydrogen atoms are chemically adsorbed on graphene surface, significant bulge and distortion of the local surface near the adsorption site present. In the surface normal, carbon atoms bonded with H/D of the monomer, ortho-dimer, meta-dimer, and para-dimer are 0.52, 0.78, 0.63 and 0.59 Å higher than the average height of the rest of carbon atoms, respectively. It can be seen that the smaller the distance between the two adsorption sites, the larger the height of the surface bulge. In order to measure the stability of different adsorption configurations, the adsorption energy $E_{ads}$ is defined as follows:

$$E_{ads} = E_{H/graphene} - \left(n \times E_{H(isolated)} + E_{graphene}\right), \qquad (1)$$

where $E_{H(isolated)}$, $E_{graphene}$ and $E_{H/graphene}$ are the total energies of an isolated hydrogen atom, the pristine graphene surface and optimized absorption system of hydrogen on graphene surface, respectively; *n* is the number of chemisorbed hydrogen atoms. The adsorption energies of hydrogen monomer, ortho-dimer, meta-dimer and para-dimer are calculated to -0.687, -2.634, -1.531, -2.627 eV, respectively. These data are in good agreement with previously reported results [16, 24, 25]. According to the adsorption energies, the bonding strength of hydrogen monomer and meta-dimer on graphene surface is close to each other, while the ortho- and para-dimers are two significantly more stable and energetically nearly degenerate configurations.

Scanning tunneling microscopy observations [13-15] have shown that monomer, ortho-dimer, and para-dimer are the three main aggregation states of hydrogen on the graphite/graphene surfaces, while meta-dimer cannot be found. Even at a low temperature of 140 K, meta-dimer cannot exist stably. This is due to the very small diffusion barrier of hydrogen atoms in the meta-dimer, allowing hydrogen atoms in the meta-dimer to reach a more stable adsorption state through surface diffusion [20]. For example, a meta-dimer can be converted into an ortho-dimer or para-dimer by diffusion of one hydrogen atom to the nearest neighboring adsorption site. Therefore, for desorption of hydrogen dimers, we only consider the ortho-dimer and para-dimer.

It has been established by previous studies [1, 17] that the formation and desorption of $H_2$ from graphene may proceed via two mechanisms: 1) The Eley−Rideal (ER) mechanism in which the chemisorbed H atom collide with an incident H and leaves the surface in the form of $H_2$; 2) the Langmuir−Hinshelwood (LH) mechanism where two physisorbed or chemisorbed H atoms combine to form $H_2$ and leave the surface simultaneously. Actually, in the absence of an incident H, effective desorption of a chemisorbed H is possible even at low-temperatures [17]. This motivates us to suggest *a third mechanism* for the desorption process of hydrogen dimers on graphene surface. Under this mechanism, desorption takes place in the form of individual H/D atoms through a two-step process: One of the H/D atoms in the dimer takes the lead in desorption to convert the dimer into a monomer, and the remained monomer leaves the surface and combines with the early-leaving H/D to form a $H_2/D_2$ molecule. For direct comparison, the LH mechanism is also referred to as *molecular mechanism*, and the two-step mechanism suggested in this work is named as *atomic mechanism*. These two mechanisms are schematically illustrated in Fig. 1. For each desorption process, the climbing image nudged elastic band (CI-NEB) method [26] is used to calculate the minimum energy path (MEP) and determine the energy barrier. Desorption of hydrogen from the graphene surface breaks the C-H/D bond. When the interactions between H/D and surface are altered, the C-H/D vibrational frequencies and phonon energies change accordingly. To obtain precise desorption barriers, it is crucial to implement zero-point energies (ZPE) correction to account for the influence of zero-point vibrations. The vibrational frequencies are calculated based on density functional perturbation theory (DFPT) [27]. The ZPE can be obtained as $ZPE = Re[\sum_{i=1}^{3N} \frac{\hbar \omega_i}{2}]$, where $3N$ and $\omega_i$ represent the number of vibration modes and the vibration angular frequency, respectively; $\hbar$ is reduced Planck constant. Only the modes with real vibrational frequencies are considered in the summation.

Figures 2(a)-2(c) show the desorption barriers of H/D in monomers, ortho-dimers and para-dimers under the atomic desorption mechanism. Desorption in the form of individual H/D atom is a typical endothermic process. The potential barriers of desorption via atomic mechanism for monomer, ortho-dimer, and para-dimer are 1.020 1.932 and 1.901 eV, respectively, which are consistent with previous DFT calculations [13]. This indicates that the formation of ortho-dimer and para-dimer significantly increases the difficulty of desorption under the atomic mechanism. After the ZPE correction, the height of desorption barrier decreases a little and the impact of ZPE on H is greater than that of D as the adsorbate.

Figures 2(d) and 2(e) present the desorption barriers of H/D in ortho-dimers and para-dimers under the molecular desorption mechanism. Compared to the dimer states, the energies of the systems drop significantly when the H/D atoms detach from the graphene surface via the molecular mechanism. The potential barriers for desorption in the form of molecules for the ortho-dimer and para-dimer are 2.227 and 1.131eV, respectively. The calculated desorption barriers are slightly lower than previous DFT data [16, 17]. This small difference may be due to whether spin polarization was included or not in calculations. From the barrier heights shown in Table 1, it can be inferred that H/D ortho-dimers and para-dimers may be more likely to desorb through the atomic and molecular desorption mechanisms, respectively. This is to be verified below.

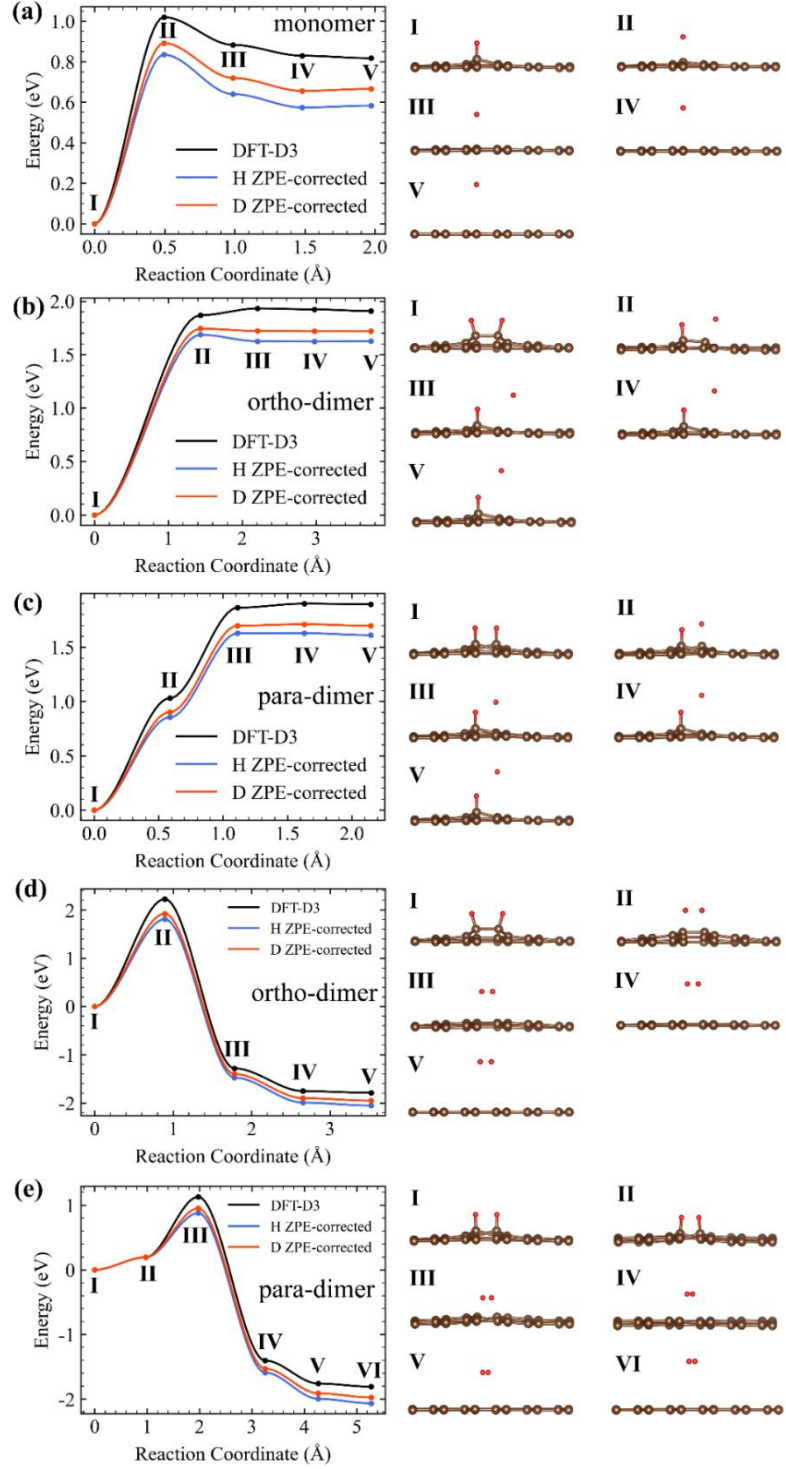

**Fig. 2.** Desorption barriers of hydrogen in monomer (a), ortho-dimer (b) and para-dimer (c) under the atomic mechanism and the molecular mechanism [ortho-dimer (d), para-dimer (e)]. The barriers with and without ZPE corrections are displayed. The side views of some intermediate configurations on the MEP are presented in the right panels. The brown and red balls represent C and H/D atoms,

respectively.

**Table 1.** The original (DFT-D3) and the ZPE-corrected barriers for monomer, ortho-dimer and para-dimer under different desorption mechanisms.

| Adsorption Configurations (desorption mechanism) | DFT-D3 (eV) | H-ZPE corrected (eV) | D-ZPE corrected (eV) |
| --- | --- | --- | --- |
| Monomer (atomic mechanism) | 1.020 | 0.835 | 0.892 |
| Ortho-dimer (atomic mechanism) | 1.932 | 1.686 | 1.743 |
| Ortho-dimer (molecular mechanism) | 2.227 | 1.810 | 1.925 |
| Para-dimer (atomic mechanism) | 1.901 | 1.629 | 1.711 |
| Para-dimer (molecular mechanism) | 1.131 | 0.879 | 0.952 |

Now we pay attention to the kinetic properties of H/D desorption from graphene surface, with focus on the role of isotope effect and quantum tunneling. A classical particle can surmount an energy barrier only if its kinetic energy exceeds the barrier height. The classical probability $P_c(T)$ corresponding to the desorption process at a given temperature $T$ can be calculated as follows [28]:

$$P_c(T) = \int_{E_b}^{\infty} p(E,T) dE = \left(1 - Erf\left[\sqrt{E_b/(k_B T)}\right]\right) + \frac{2}{\sqrt{\pi}} \sqrt{E_b/(k_B T)} e^{-\left(\frac{E_b}{k_B T}\right)}, \quad (2)$$

where $p(E,T) = 2\pi \left(\frac{1}{\pi k_B T}\right)^{\frac{3}{2}} \sqrt{E} e^{-\frac{E}{k_B T}}$ is the particle's kinetic energy distribution [29] in a thermal equilibration system; $E$ is the particle's kinetic energy; $k_B$ and $E_b$ are the Boltzmann constant and the desorption barrier height, respectively.

When the H/D atoms are treated as quantum particles, the transmission probability $P_Q(T)$ is calculated as follows:

$$P_Q(T) = \int_{E_0}^{\infty} p(E,T) T_r(E) dE \cong \int_{E_0}^{E_m} p(E,T) T_r(E) dE, \quad (3)$$

where $T_r(E)$ is the transmission coefficient of the particle with kinetic energy $E$ calculated by the transfer matrix (TM) method (The details of the TM method are provided in the Supporting Materials). When we calculate the transmission coefficient using the TM method, the particle mass is that of one H/D atom and one $H_2/D_2$ molecule, respectively, for the desorption process under the atomic and molecular mechanism. The integral lower limit ($E_0$) is the energy of initial and final state for exothermic and endothermic reactions, respectively. In practice, the integration upper limit ($E_m$) is set to 10 eV to ensure that the results converge to the desired precision at an energy sampling interval of $10^{-4}$ eV.

The probabilities of H desorption from graphene surface in the form of H atoms (atomic mechanism) and $H_2$ molecules (molecular mechanism) at a given temperature $T$ are labeled as $P|_H(T)$ and $P|_{H_2}(T)$, respectively. For dimer desorption via the molecular mechanism, $P|_{H_2}(T)$ is the probability of two H atoms simultaneously crossing the desorption barriers shown in Figs. 2(d) and 2(e). For hydrogen monomer, $P|_H(T)$ is the probability of a single H atom crossing the desorption barrier shown in Fig. 2(a). For dimer desorption via the atomic mechanism, the first H atom desorbs with the probability $P|_{dimer-1st\ H}(T)$, which is the probability of one H atom in the dimer crossing the desorption barriers shown in Figs. 2(b) and 2(c). When one of the H atoms in the dimer desorbs, the dimer is converted into a monomer. Since desorption of the second H of a dimer depends on desorption of the first H, the desorption probability of the second H is as follows:

$$P|_{dimer-2nd\ H}(T) = P|_{dimer-1st\ H}(T) \times P|_{monomer-H}(T). \quad (4)$$

Suppose that there are $n$ H-dimers (e.g., ortho-dimer) on the surface. At temperature $T$, the desorption probability of the first and second H is $P_1(T)$ and $P_2(T)$, respectively. Then the number of desorbed $H_2$ is $N_{d,H_2}(T) = (nP_1(T) + nP_2(T))/2$, and the total desorption probability is $P_{d,tot}(T) = \dfrac{N_{d,H_2}(T)}{n} = \dfrac{nP_1(T) + nP_2(T)}{2n}$, which is reduced to $P_{d,tot}(T) = \dfrac{P_1(T) + P_2(T)}{2}$. Therefore, the total probability of H-dimer desorption through the atomic mechanism, $P|_{dimer-H}(T)$, is as follows:

$$P|_{dimer-H}(T) = \dfrac{P|_{dimer-1st\,H}(T) + P|_{dimer-2nd\,H}(T)}{2}. \tag{5}$$

Similar analysis apply to the desorption process of D-dimers.

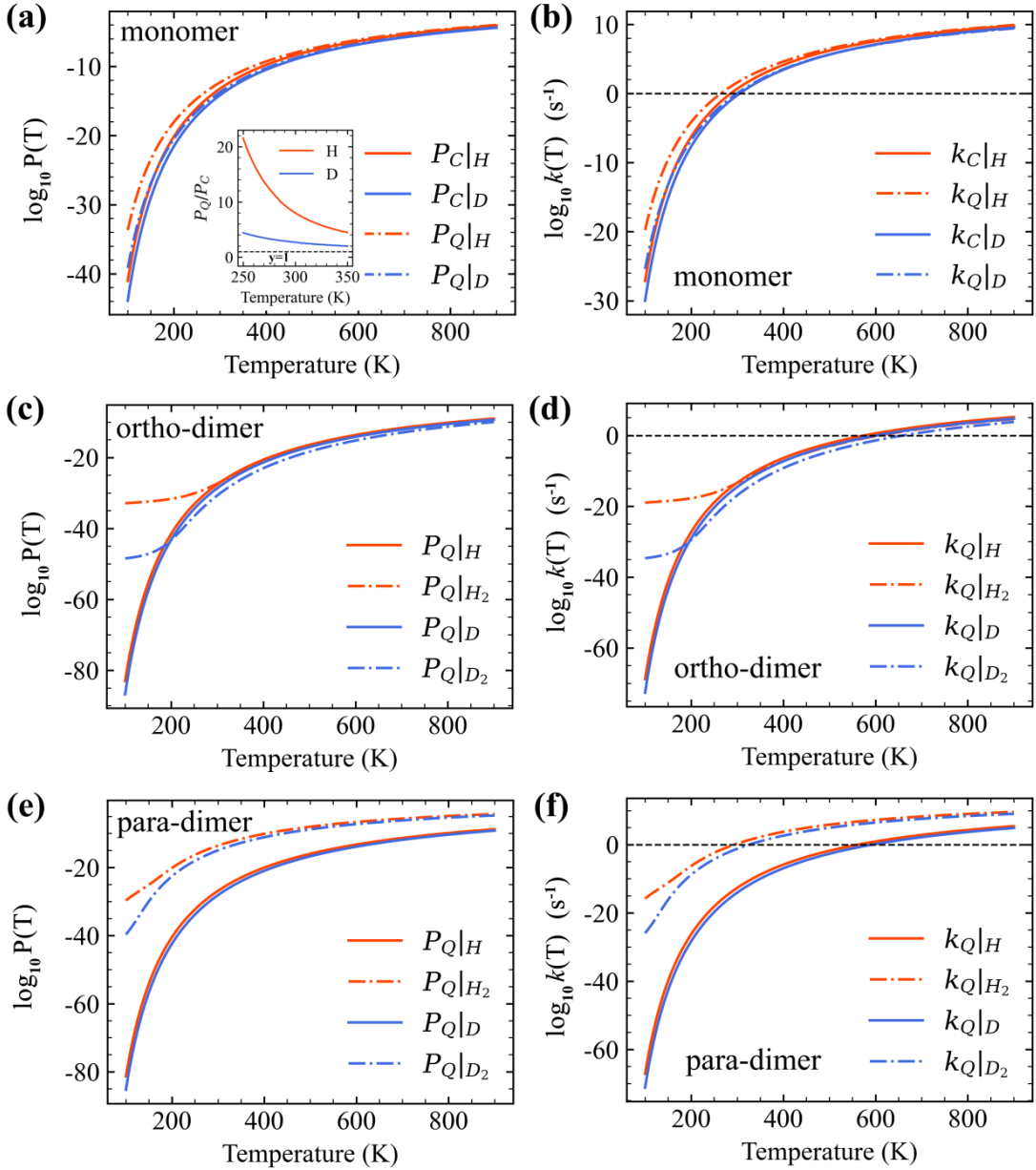

**Fig. 3.** The probabilities and the rate constants of hydrogen desorption from the graphene surface under different desorption mechanisms with H and D as adsorbates.

Figures 3(a), 3(c) and 3(e) show the probability of desorption from graphene surface under different desorption mechanisms with H and D as adsorbates. In the case of monomer desorption [Figs. 3(a)-(b)], it is clearly seen that quantum tunneling plays a significant role in enhancing the barrier-crossing probability and the rate constant at room temperature and below, for which the quantum probability (and rate constant) is much larger than the classical one. For temperatures below 200 K, most particles possess very low kinetic energy. Under the molecular mechanism of dimer desorption, quantum tunneling enables particles with kinetic energies in the range (0, $E_b$) to have nontrivial transmission probability of crossing the barrier. In the low temperature region, quantum tunneling plays a dominant role in hydrogen desorption, resulting in a significantly higher quantum probability with comparison to the classical one [Figs. S2(c) and S2(d)]. Due to its smaller mass with comparison to D atom, much more significant effects of quantum tunneling are found for H. Therefore, at the same temperature, the lower desorption barrier and more significant tunneling effects result in a higher desorption probability of H. As the temperature increases, the kinetic energies of the particles increase on average, causing a decrease in the proportion of particles in the low-energy region and an increase in the proportion of particles with energies above than the barrier height. As a result, the classical probability tends to converge with the quantum probability due to the diminishing tunneling effects.

As seen from Figs. S2(a) and S2(b), for desorption of the first H/D in dimers, the effects of quantum tunneling are less pronounced under the atomic mechanism compared to the molecular mechanism. This is due to the fact that, in the endothermic reaction processes, quantum tunneling only affects particles with kinetic energies in the interval of [$E_0$, $E_b$), and particles with kinetic energy less than $E_0$ cannot penetrate the barrier through quantum tunneling and decay exponentially as evanescent waves [23]. At reasonable temperatures, the kinetic energy of most particles is less than $E_0$ according to the kinetic energy distribution function. Therefore, quantum tunneling has minor effects under the atomic mechanism for dimer desorption.

Figures 3(c) and 3(e) compare the quantum desorption probabilities of hydrogen dimers under two different mechanisms. For the para-dimers, the desorption probability under the molecular mechanism is always larger than that under the atomic mechanism across the entire temperature range. In the case of ortho-dimers, the desorption probability of the molecular mechanism surpasses that of the atomic mechanism when the temperature is lower than 300 K. This is attributed to the significant promotion of desorption in the form of hydrogen molecules by quantum tunneling at low temperatures. As the temperature increases, the influence of quantum tunneling diminishes. At temperatures above 300 K, the atomic mechanism begins to play a significant role in the desorption process of H ortho-dimers. In the case of D ortho-dimers, a similar transition in dominant desorption mechanism occurs at ~ 200 K.

For desorption of hydrogen monomers, the reaction rate constant $k$ at a given temperature $T$ can be calculated as follows [19]:

$$k_H(T)|_{monomer} = v_{monomer} \times P_Q|_H(T), \quad (6)$$

where $v$ is the attempting frequency factor, which approximately takes the frequency of perpendicular vibrational modes of the adsorbed atom. Desorption of dimers under the atomic mechanism is a two-step process. The rate constant for which one hydrogen atom in the dimer desorbs firstly is as follows:

$$k_H(T)|_{dimer-1st\ H} = v_{dimer} \times P_Q|_H(T), \quad (7)$$

Desorption of the second hydrogen atom is actually the desorption process of a hydrogen monomer. Therefore, the rate constant for desorption of the second hydrogen atom is: $k_H(T)|_{dimer-2nd\ H} = k_H(T)|_{monomer}$.

The total rate constant for dimer desorption under the atomic mechanism can be obtained as follows [30]:

$$\frac{1}{k_H(T)|_{dimer}} = \frac{1}{k_H(T)|_{dimer-1st\ H}} + \frac{1}{k_H(T)|_{dimer-2nd\ H}} \ . \tag{8}$$

The rate constant under the molecular mechanism is $k_{H_2}(T) = v_{dimer} \times P_Q|_{H_2}(T)$.

Table 2 lists the frequencies of perpendicular vibrational modes H (D) derived from DFPT calculations for hydrogen monomer, ortho-dimer, and para-dimer. The calculated vibrational frequencies of H and D monomers compare well with the experimental data measured on the graphite surface using high-resolution electron energy loss spectroscopy, with a difference of less than 2% [31, 32].

**Table 2.** Vibrational frequencies and energies of the perpendicular modes of H (D) monomer, ortho-dimer and para-dimer derived from DFPT calculations.

|  | Frequency (THz) | Energy (meV) |
|---|---|---|
| monomer (H) | 80.94 | 334.75 |
| monomer (D) | 59.24 | 245.01 |
| ortho-dimer (H) | 86.18 | 356.41 |
| ortho-dimer (D) | 63.19 | 261.33 |
| para-dimer (H) | 82.57 | 341.48 |
| para-dimer (D) | 60.48 | 250.11 |

According to previous thermal desorption spectrum (TDS) experiments [15, 16, 33], the deuterium-adsorbent graphite/graphene surface has obvious $D_2$ desorption peak in the temperature range of 350-650 K when heated with a rate of 1-2 K/s. When the deuterium coverage is very low (less than $1 \times 10^{15}$ cm$^{-2}$), a single desorption peak is divided into two peaks which locate in the temperature range of 400-520 K and 580-620 K, respectively. The shape and position of the desorption peak are closely related to the deuterium coverage on the surface. Figures 3(b), 3(d) and 3(f) show the reaction rate constant of hydrogen desorption in different aggregation states. At the same temperature, the rate constant of H as the adsorbate is larger than that of D as the adsorbate. When the rate constant reaches 1 s$^{-1}$ (matching the heating rate of 1-2 K/s in TDS experiments), hydrogen desorption from the graphite surface should be observable in the TDS experiment. For monomers with D as the adsorbate, the temperature at which the rate constant reaches 1 s$^{-1}$ is 300 K. For ortho-dimers with D as the adsorbate, the critical temperatures for desorption under atomic and molecular mechanisms are 602 K and 659 K, respectively. For para-dimers with D as the adsorbate, the critical temperatures for desorption through atomic and molecular mechanisms are 591 K and 325 K, respectively. According to our calculations, desorption of D in para-dimers and ortho-dimers should begin at about 330 K and 600 K, respectively. The calculated onset temperatures for desorption of ortho-dimers are in good agreement with the second desorption peak observed in the TDS experiments [15, 16, 33]. The desorption temperature of D in para-dimers is slightly lower than the onset temperature of the first TDS peak. Our calculation results agree with TDS experiments and indicate that the two desorption peaks in TDS experiments actually correspond to desorption of D in para-dimers and ortho-dimers, respectively. STM experiments of graphene surface with a hydrogen coverage of 0.15% [14] show that, compared to the data measured at 140 K, the proportion of monomers on the graphene surface at room temperature decreases significantly down to less than 1%. The proportion of para-dimers on graphene surface at 140 K and room temperature were determined to be ~ 85% and 73%, respectively. These observations indicate that at room temperature, effective desorption of H in monomers has occurred and hydrogen in para-dimers start to desorb by a small amount. The experimental observations are also consistent with our calculations.

Using the rate constant, the proportion of the hydrogen desorption amount under the two different mechanisms can be obtained as follows:

$$\eta_H(T) = \frac{k_H(T) \times t}{k_H(T) \times t + k_{H_2}(T) \times t} = \frac{k_H(T)}{k_H(T) + k_{H_2}(T)}, \qquad (9)$$

$$\eta_{H_2}(T) = \frac{k_{H_2}(T) \times t}{k_H(T) \times t + k_{H_2}(T) \times t} = \frac{k_{H_2}(T)}{k_H(T) + k_{H_2}(T)}, \qquad (10)$$

where $\eta_H(T)$ and $\eta_{H_2}(T)$ are the desorption proportions for the atomic and molecular mechanism at temperature $T$, respectively; $t$ is the reaction time. The results are shown in Fig. 4.

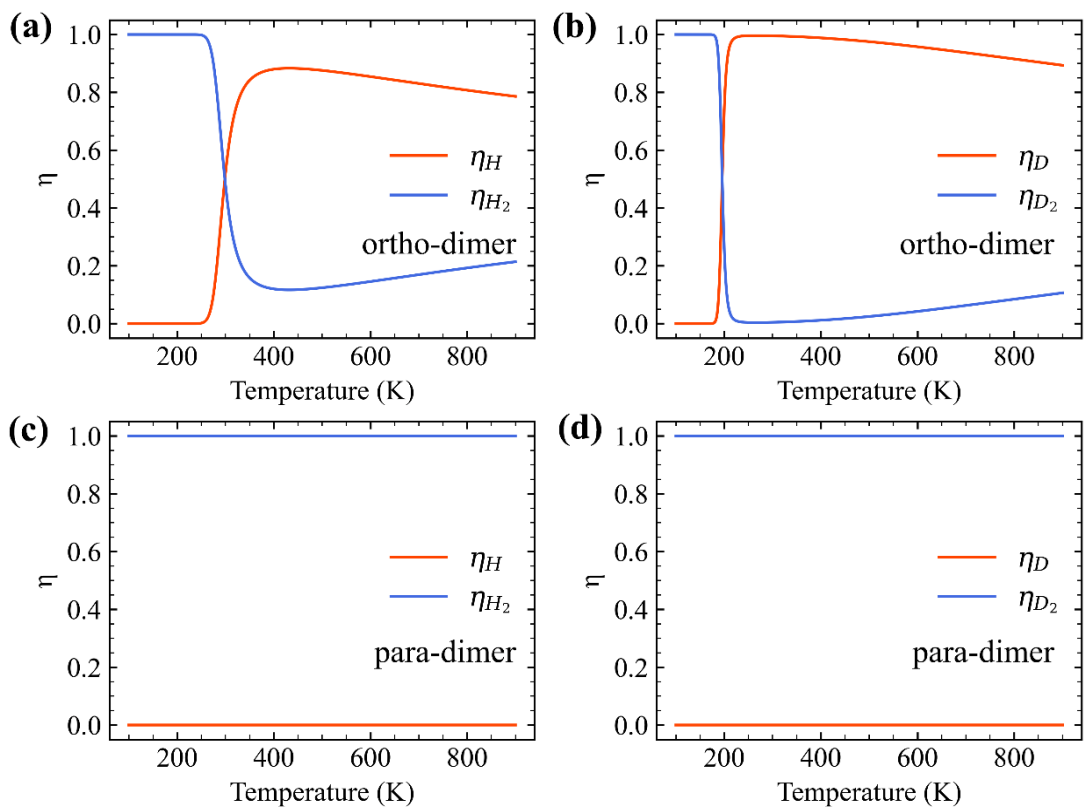

**Fig. 4.** The proportion of the hydrogen desorption amount under the two different desorption mechanisms in ortho-dimers and para-dimers.

For ortho-dimers with H and D as adsorbates [Figs. 4(a)-(b)], the dominant desorption mechanism switches at the temperature of ~ 300 K and 200 K, respectively. It is worth noting that although the molecular mechanism does not prevail in the high-temperature region, its role slowly increases as the temperature rises and remains nontrivial. By contrast, it is clearly seen that the molecular mechanism is always dominant in desorption of para-dimers [Figs. 4(c)-(d)]. In para-dimers, desorption under the atomic mechanism is almost completely suppressed. The difference of dominant desorption mechanism between the ortho- and para-dimers can be understood from the MEP (Fig. 2). Under the atomic mechanism, the energy barriers for desorption of the first H/D are similar for the ortho- and para-dimers, and exactly the same for desorption of the second H/D as monomers. However, under the molecular mechanism, the energy barriers for desorption of $H_2/D_2$ differ largely for the ortho- and para-dimers [see Figs. 2(d)-(e) and Table 1]. The much lower energy barrier leads to dominant molecular mechanism for the para-dimers desorption in all temperature region.

To summarize, this work systematically studies desorption of hydrogen from graphene surfaces based on first-principles calculations, in combination with the TM method to include the effects of quantum tunneling. At low temperatures, it is shown that quantum tunneling significantly enhances the desorption process of hydrogen monomers and dimers. With the increase of temperature, the influence of quantum tunneling is gradually reduced. The adsorption energies and desorption barriers indicate that for both ortho-dimers and para-dimers, the presence of neighboring hydrogen atoms enhances the binding strength of chemisorbed hydrogen atoms to the graphene surface. For dimer desorption, two microscopic mechanisms are considered: The traditional molecular form of desorption (molecular mechanism), and a newly suggested mechanism, namely, the atomic mechanism, which proceeds via a two-step process. In the low-temperature region, the molecular mechanism dominates the desorption process of all types of dimers. While the molecular mechanism continues to play a major role in the desorption process of para-dimers, the atomic mechanism starts to prevail in the high-temperature desorption of ortho-dimers. The reason for the transition of dominant desorption mechanism is that the contribution of quantum tunneling to desorption under the molecular mechanism gradually decreases with the increase of temperature. The transition of the dominant desorption mechanism of ortho-dimers with H and D as adsorbates takes place at $T \sim 300$ K and 200 K, respectively. The results are expected to be tested by future experimental measurements.

**Acknowledgements**

This work is financially supported by the National Natural Science Foundation of China (No. 12074382, 11474285). We are grateful to the staff of the Hefei Branch of Supercomputing Center of Chinese Academy of Sciences, and the Hefei Advanced Computing Center for support of supercomputing facilities.

**Supplemental Materials for** "Quantum Tunneling Enhanced Hydrogen Desorption from Graphene Surface: Atomic versus Molecular Mechanism"

Yangwu Tong[1,2] and Yong Yang[1,2]*

*1. Science Island Branch of Graduate School, University of Science and Technology of China, Hefei 230026, China.*

*2. Key Lab of Photovoltaic and Energy Conservation Materials, Institute of Solid State Physics, HFIPS, Chinese Academy of Sciences, Hefei 230031, China.*

*Corresponding Author: yyanglab@issp.ac.cn

**Theoretical methods**

In this study, all density functional theory (DFT) calculations considering spin polarization are carried out using the Vienna *ab initio* simulation package (VASP) [1, 2]. The electron-electron exchange correlations and electron-ion interactions are described using Perdew-Burke-Ernzerhof (PBE) type generalized gradient approximation (GGA) [3] and projector augmented wave (PAW) [4, 5] method, respectively. DFT-D3 dispersion correction is carried out to deal with van der Waals interactions [6]. The plane waves with energy-cut of 600 eV are employed for the description of electron wave functions. A $4 \times 4 \times 1$ Monkhorst-Pack k-mesh [7] is chosen for *k*-point sampling of the Brillouin zone. The graphene surface is modeled by a (5×5) supercell with a vacuum space of ~ 15 Å in the z-direction that repeats periodically along the xy plane. The climbing image nudged elastic band (CI-NEB) method is used to obtain the potential barriers of the minimum energy path of hydrogen desorption from the graphene surface [8]. During the structural relaxation, the unit cells are optimized to guarantee that the total energy and forces at all atomic sites converge to within a threshold of $10^{-4}$ eV and 0.05 eV/Å, respectively. The vibrational modes are calculated based on density functional perturbation theory (DFPT) [9]. The barriers are then corrected by zero-point energies (ZPE), which are derived from DFPT calculations.

The transfer matrix (TM) method, a highly accurate computational approach for calculating transmission probabilities of quantum particles pass through a potential field of arbitrary shape, is employed to investigate the quantum effects of hydrogen desorption. The TM method regards the process of quantum particles passing through potential barriers as the process of particles successively passing through multiple potential barrier slices. In order to calculate the transmission coefficient of hydrogen as a quantum particle across the desorption barrier using the TM method, the desorption barrier is divided into a series of slice chains ($S_1$, $S_2$, ..., $S_j$, ..., $S_{n-1}$, $S_n$). When the slice width is small enough, each slice can be approximated as a trapezoidal potential barrier and corresponds to a coefficient matrix. The transfer matrix $M$ of the whole process is the chain product of all coefficient matrices [10], which can be obtained as follows:

$$M = M_n M_{n-1} \cdots M_j \cdots M_2 M_1 = \begin{bmatrix} m_{11} & m_{12} \\ m_{21} & m_{22} \end{bmatrix}, \quad (S1)$$

where $M_j$ is the coefficient matrix corresponding to the $j$th slice $S_j$. The incident wave function and outgoing wave function of the particle can be related by the transfer matrix as follows [11]:

$$\begin{bmatrix} A_R \\ K_R \end{bmatrix} = M \begin{bmatrix} A_L \\ K_L \end{bmatrix} = \begin{bmatrix} m_{11} & m_{12} \\ m_{21} & m_{22} \end{bmatrix} \begin{bmatrix} A_L \\ K_L \end{bmatrix}, \quad (S2)$$

where $A_L$ and $A_R$ are the incident amplitude and the outgoing amplitude; $K_L$ and $K_R$ are the incident wave vector and the outgoing wave vector. The transmission coefficient $T_r(E)$ is calculated by [10]

$$T_r(E) = \left|\frac{A_R}{A_L}\right|^2 \times \frac{K_R}{K_L} = \frac{|M|^2}{|m_{22}|^2} \times \frac{K_R}{K_L}, \quad (S3)$$

where $|M|$ is the determinant of $M$.

**Mathematical definition of reaction coordinate in Fig. 2.**

In this study, the reaction coordinate reflects the change of all atomic sites in three-dimensional space during desorption, which is defined mathematically as follows:

$$C_k = \sum_{j=2}^{k} \sum_{i=1}^{N} \left|\vec{R}_{i,j} - \vec{R}_{i,j-1}\right|, \quad (k \geq 2) \quad (S4)$$

where $C_k$ is the reaction coordinate for the $k$th state, $\vec{R}_{i,j}$ is the coordinate vector of the $i$th atom in the $j$th state, $N$ is the number of atoms in the unit cell. The reaction coordinate of the initial state $C_1$ is zero.

**Structures of hydrogen monomer and three typical hydrogen dimer structures on the graphene surface**

Through structural relaxation, we obtain the structural models of hydrogen monomer and three types of hydrogen dimers in which two hydrogen atoms are chemisorbed on the same hexagonal ring of carbon atoms. The top and side views of these structures are shown in Fig. S1 with the aid of VESTA [12].

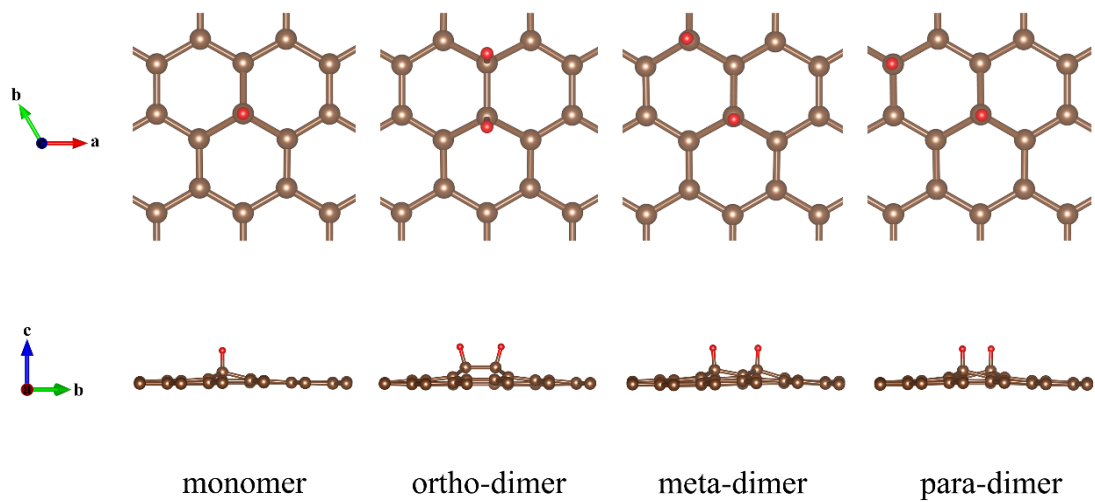

monomer    ortho-dimer    meta-dimer    para-dimer

**Fig. S1.** Top and side views of hydrogen monomer and three hydrogen dimers on the graphene surface. The brown and red balls represent carbon and hydrogen atoms, respectively.

**Comparation of classical and quantum probabilities of hydrogen desorption from the graphene surface in dimers**

We compare the classical and quantum probabilities of hydrogen desorption from the graphene surface under different desorption mechanisms in dimers with H and D as adsorbates. For the desorption in the form of H atoms in the dimer, the probability of desorption of the first H atom $P|_{dimer-1st\,H}(T)$ is the probability of one H atom in the dimer crossing the desorption barrier shown in Figs. S2(a) and

S2(b). For the desorption of hydrogen in the form of $H_2$ molecules in the dimer, $P|_{H_2}(T)$ is the probability of two hydrogen atoms crossing the desorption barrier shown in Figs. S2(c) and S2(d).

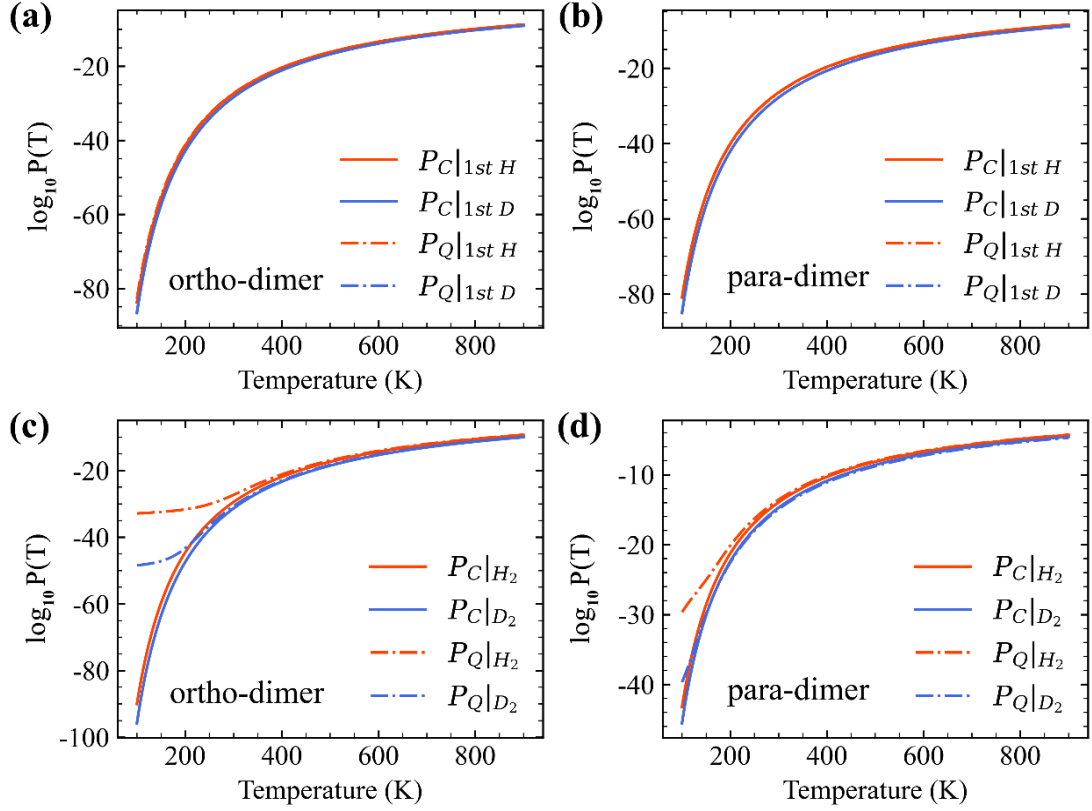

**Fig. S2.** Comparation of classical and quantum probabilities of hydrogen desorption from the graphene surface under different desorption mechanisms in dimers with H and D as adsorbates.